# A Proposal for Outlier and Noise Detection in Public Officials' Affidavits

Rodrigo López-Pablos[1,2] and Horacio D. Kuna[3]

[1]Escuela de Posgrado, Facultad Regional Buenos Aires,
Universidad Tecnológica Nacional, Argentina
[2]Санкт-Петербургский Национальный Исследовательский Университет
Информационных Технологий, Механики и Оптики, Россия
[3]Departmento de Cs. de la Computación, Facultad de Ciencias Exactas, Químicas y
Naturales, Universidad Nacional de Misiones, Posadas, Argentina
{rodrigo.lopezpablos,hkuna}@gmail.com

**Abstract.** Outlier and noise detection processes are highly useful in the quality assessment of any kind of database. Such processes may have novel civic and public applications in the detection of anomalies in public data. The purpose of this work is to explore the possibilities of experimentation with, validation and application of hybrid outlier and noise detection procedures in public officials' affidavit systems currently available in Argentina.

## 1 Introduction

Data mining and information exploitation processes have been scarcely used for solving civic issues related to the public sector, and outlier and noise detection processes are not the exception. Public data can be subject to anomalies and noise, like any kind of data in any given database (DB). However, the implications of the discovery of corrupt behavior in the public data of public officials and in the quality of such data might have significant and profound effects on society's welfare since corruption affects the social fabric and the quality of life of its populations. In this context, data mining might be a particularly useful tool for those citizens and civil organizations fighting against corruption since it can aid in the discovery of corrupt social fabric, thereby highlighting the usefulness of data mining as a tool in social sciences, monitoring and scientific research [1].

This paper is organized as follows: Subsection 1.1 presents the hypothetical questions posed in this work; Section 2 discusses the state of the art in data mining techniques aimed at detecting corrupt behavior; Section 3 analyzes open affidavit systems;

Section 4 deals with the proposed outlier detection procedures; Section 5 addresses the experimentation with such procedures; and Section 6 presents the conclusions of this work.

### 1.1. Research Questions

This research proposal poses the following questions:

- Can the experimentation with, application and use of outlier detection techniques and processes in public databases of official affidavits be considered a useful tool to find signs of corrupt behavior in public administration and in the fight against corruption?
- Is it possible to assess the quality of such databases through noise and outlier analysis?
- Which outlier and noise detection procedures can be applied considering the nature of such systems?

## 2. State of the Art

Data mining and information exploitation techniques and processes have been scarcely used or even completely disregarded as an aiding civic tool in the fight against corruption in public offices. However, there is relevant research in the literature concerning mining techniques aimed at detecting financial and accounting fraud or corruption in the private sector.

Several authors [2,3] have provided a comprehensive classification of data mining use against private corruption. Such corruption cases might take the form of financial and accounting fraud, including banking fraud, credit card fraud, money laundering and mortgage fraud, cases of complex private fraud, and complex financial fraud involving the forging of corporate information, financial speculation, etc. A classification of techniques used in the field is presented below.

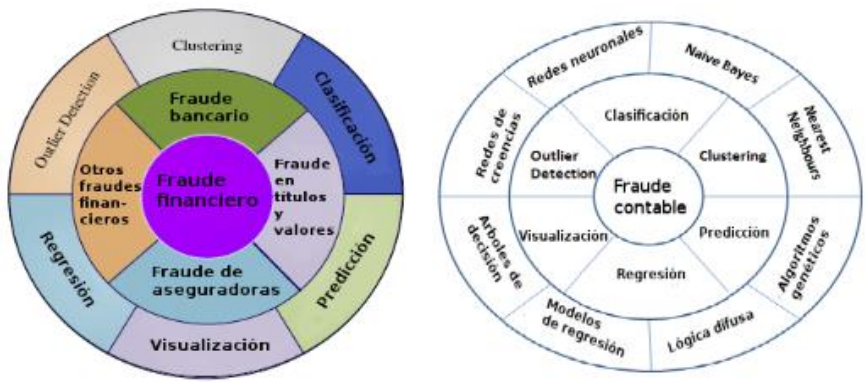

**Fig. 1.** On the left, a classification of information exploitation techniques for financial fraud detection [2]. On the right, data mining techniques and algorithms used for the detection of accounting or fiscal fraud [3].

Figure 1 shows that the use of these tools has been under-exploited in the public sector both for civic use and against corruption since there are no previous relevant works in the literature revealing any interest in the problem of public corruption.

In addition, without detriment to other data mining techniques and information exploitation processes, techniques for the detection of outliers or anomaly values have not yet been implemented to help solve fraud scenarios, let alone in the area of corruption in the public sector.

## 3. Open Affidavit Systems as an Alternative Solution in the Fight Against Public Corruption

In Argentina, affidavit systems are information systems regulated by affidavit regimes or systems that have three basic functions for which they were implemented [4]:

[i]    Controlling the patrimonial evolution of public officials in order to prevent illicit gain and other corruption-related crimes.

[ii]    Detecting and preventing conflicts of interest and incompatibilities with public office.

[iii]    As a mechanism for transparency and prevention of public corruption.

In terms of prevention, all public official affidavit regimes constitute a tool that enables: control of appropriate compliance with public office duties by public officials, prevention of deviation from their ethical duties and correction of any detected non-compliance.

### 3.1 Open Affidavit Application Software and Systems

From the exploratory collection of public data in Argentina, the Open Financial Affidavit interactive system [5] jointly developed by Diario La Nación and the NGOs Directorio Legislativo, Poder Ciudadano and Asociación Civil por la Libertad y la Justicia, in force since 2013, was selected for the experiment. Out of 1550 financial affidavits on the interactive site, 539 affidavits corresponded to 99 public officials of the executive branch of government, 843 affidavits corresponded to 313 officials of the legislative branch, and 168 corresponded to 87 judicial branch officials, according to the updated version as of April 14, 2015. The data preparation strategy to perform the experiment consisted in only considering public officials' real estate.

The open affidavid database has the following attribute structure for each public official's affidavit:

```
dj.funcionario =  (ddjj_id, ano, tipo_ddjj, poder, persona_id, nombre,
                   ingreso, cargo, jurisdiccion, cant_acciones,
                   descripcion_del_bien, destino, localidad, nombre_bien_s,
                   origen, pais, porcentaje, provincia, tipo_bien_s,
                   titular_dominio, vinculo, superficiem2, val_decl,
                   valor_patrim)
```

The database created for the experiment has a total of 6627 tuples with 24 attributes since, from the original database with 39 initial attributes, 13 were discarded when partially or completely empty fields or nominal inconsistences in data imputation were found. In addition, some attributes became redundant after the homogenization and standardization of the three (3) generated attributes: `dj.patrimoniales (Gen)=(`**`superficiem2, valor_patrim, val_decl`**`)`.

In the experiment conducted, all real estate appraised in foreign currency was converted to Argentine pesos and inflation-adjusted (`valor_patrim`), the real estate area was homogenized to square meters (`superficiem2`) and the real estate value declared by the public official (`val_decl`) was classified according to its relative fiscal valuation into fiscal, sub-fiscal, market value, or no value declared.

## 4. Proposed Hybrid Outlier and Noise Detection Procedures

Anomaly fields are defined as data so different from the other data belonging to the same data set [6], *i. e.* a database containing such fields, that it can be deemed created by a different mechanism. It is precisely the discovery of such mechanisms that the analysis of each database pursues.

Hybrid detection methods –which combine various algorithms from different learning approaches– have recently come to be regarded as processes, and the combination of different procedures allows outlier detection with a confidence level of over 60% [7]. Hybrid outlier detection methods offer the advantage of combining different techniques and algorithms to achieve the same goal. These methods include: LOF (Local Outlier Factor) and metadata, or LOF and K-Means, for numerical DBs as well as induction algorithms such as C4.5, PRISM, Information Theory, and Bayesian Networks; and clustering algorithms such as LOF, DBSCAN and K-Means for alphanumeric DBs with both supervised and unsupervised procedures.

Table 2 below shows the use of hybrid methods with different approaches, depending on the learning approach of the algorithms involved, that is, supervised or unsupervised learning aimed at outlier and noise detection. Hybrid methods are the best alternative to obtain the highest amount of information, reduction of search space and process optimization [8, 9, 10]. Recent research has shown that the combination of different types of algorithms as well as the combination of procedures optimizes outlier discovery [7]. In line with such paradigm, the following two alphanumeric procedures are proposed to aid civil auditors:

**Table 1**. Hybrid procedures proposed according to environment, algorithm and approach [7]

| Hybrid procedures | Environment | Algorithms and techniques | Approaches |
|---|---|---|---|
| I | Alphanumeric DBs with a target attribute | C4.5, Information theory; LOF | unsupervised; supervised |
| II | Alphanumeric DBs with no target attribute | LOF; DBSCAN; C4.5; RB; PRISM; K-Means | unsupervised; supervised |

As it can be observed in hybrid procedures I and II, procedure I detects outlier fields in alphanumeric databases containing a class attribute [11,12,7] while procedure II also detects fields in alphanumeric bases, but with no target attribute [13,7].

Since affidavits are normally comprised of alphanumeric data, the potentiality of applying hybrid procedures I and II (table 2), corresponding to the hybrid procedures for the detection of anomaly data recently developed by [7], becomes evident. Such procedures are considered suitable due to the following reasons:

[i]   They were developed recently and represent the state of the art regarding the detection of anomaly fields and noise.

[ii]  They are optimal procedures for the detection of noise in alphanumeric databases, which are the type of database in which public data are generally stored.

[iii] They are easily applicable and executable with currently available data mining software.

The suitability of these hybrid procedures makes them a possible solution in civil auditing for the improvement of public and civic data of a given population. A possible solution and hypothetical application of civil auditing on open affidavit databases based on the above mentioned alphanumeric procedures are presented below:

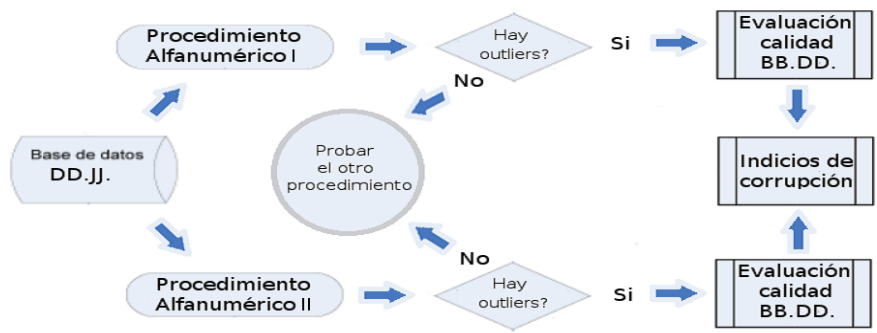

**Fig. 2.** Proposal for the application of hybrid procedures for the detection of anomaly fields in alphanumeric affidavit databases [Our own design].

The proposed procedure (Figure 4) involves the application of hybrid alphanumeric anomaly detection procedures to detect outliers in the prepared affidavit databases. The possibility of detecting false positives proposes a circular feedback in both alphanumeric procedures proposed since the rest of the procedure is the same regardless of the presence or absence of outliers. Thus, the aim is to assess the quality of the public data involved in a preliminary analysis and, in a more profound *ex post facto* analysis based on detected fields and attributes, to assess the signs of implicit corrupt behavior as output of the public information processed.

# 5. Experimentation and Discussion of the Proposed Methodology

The experiment stage is followed by the validation of the system proposed in Figure 4 in public affidavits. Since the validity of tuples and attributes of real property affidavit databases cannot be determined, suspicion of anomalies is associated to certain characteristic economic parameters of the real property which are not trivial to its composition as well as to the quality of the collected data.

## 5.1 Validation of the Hybrid Procedures Used in Affidavit Databases

From the experimental implementation of procedure I in the unsupervised learning stage, considering the declared value as the class attribute of the C4.5 induction algorithm, significant attributes are obtained, thus acquiring the greatest amount of information (table 3). Six input-output bins were designed on such attributes thus simulating an information system. Then, the mining flows with the LOF algorithm were executed *ex post facto*, where "∞" is the number of tuples which are suspected of containing anomalies.

**Table 2.** Input-output bins with detected outliers.

| Input Bins – (Output) | Detected outliers | Suspicious anomaly bins Mean or Mode (Mode) |
|---|---|---|
| `superficiem2(val_decl)` | 968 (∞) | 38733.15 (not declared) |
| `ano(val_decl)` | 122 (∞) | 2001 (Market) |
| `nombre_bien_s(val_decl)` | 209 (∞) | Horizontal Prop. (Fiscal) |
| `porcentaje(val_decl)` | 252 (∞) | 29.18528 (no data) |
| `val_patrim(val_decl)` | 1130 (∞) | 146528.3 (Subfiscal) |
| `vinculo(val_decl)` | 30 (∞) | Co-inhabitant (Subfiscal) |

Where suspicious anomaly bins were associated to the following mean values and modes: [i] non-declared real property of an average area of $38733m^2$, [ii] 2001 affidavits of real estate declared at its market value, [iii] horizontal property at fiscal value, [iv] average shareholding percentage of 29.19% with no monetary valuation, [v] average real estate value of ARS\$146,528 at subfiscal value, and [vi] subfiscally valued properties of the co-inhabitant. According to the information theory [14], the real estate value and the area seem to create more noise and entropy in the affidavit system, quantitatively speaking.

In the execution of procedure II without a class attribute, following the determination rules of outlier [7] in the first phase, LOF-DBSCAN and the combination of classification algorithms C4.5-RB-PRISM were applied. As a result, 2531 tuples suspected of containing anomalies were detected; *i. e.*, 38.19% of 6627 tuples. An outlier database is designed for the following phase which is transformed ex post facto to apply K-Means in two clusterings, obtaining the following results:

**Table 3.** Distance from centroid for each attribute

| Atribute(id)<br>(Cluster_0) | Distance value<br>(Cluster_0) | Mean value<br>(Cluster_1) |
|---|---|---|
| ddjj_id(1) | **1.841** | 1.079 |
| ano(2) | **1.518** | 1.079 |
| superficiem2(22) | **2.033** | 1.079 |

Upon executing the K-Means algorithm, it was observed that the furthest centroid contained the real estate area, year and affidavit identification attributes; where the former – superficiem2 – is the furthest attribute as well as the most suspicious of containing anomaly fields.

### 5.2 Discussion of Results of the Experimentation

The anomaly noise produced by the area, valuation, name of the property, shareholding percentage, year and relationship attributes (table 3) foresees an information system that shows the irregularities committed by public officials when appraising and declaring their properties to the citizenship. On the other hand, the Area attribute shows an important source of dispersion that could reveal signs of extremely large properties together with other properties which are too small to be considered as such.

From the first induction analysis with a class attribute –procedure I–, through C4.5 algorithm, the following rules were observed:

[i]   Properties less than 6500$m^2$ tend to be declared at subfiscal value.

[ii]  properties larger than 6500$m^2$ – between mid 2005 and 2012 – tend to be not declared, just as garages, lands, plots and horizontal properties without specification greater than 6500$m^2$ and before 2005.

[iii] when the official owns a house and a share percentage higher than 46.3% [100%; 37.9%) or lower than 37.9 (37.9% ; 0%] they will prefer not to declare its valuation, but will tend to declare it subfiscally when they hold a share of between 46% and 38% (46.3%; 37.9%).

[iv]  If the public official has an apartment, between mid 2003 and mid 2005, with an area less than or equal to 26$m^2$, it tends to be declared at its fiscal value,

while it is not declared if it is prior to 2003. However, apartments larger than 26m$^2$ are simply not declared.

[v] Public officials' stores with an appraised value greater than ARS$76,493 [∞, $76,493) or less than $10,615 ($10,615, 0] tend to be declared at their market value; except for the stores whose value is between [ARS$76,493 and ARS$10,615]. In such case, they will only be declared at their market value provided that the real property is registered under the name of the official's spouse and not under the official's name, in which case they will tend not to declare it.

## 6. Conclusions and Future Work

In this work, outlier and noise detection hybrid processes were developed and combined to be used in alphanumeric databases of affidavits of real property. It constitutes original work dealing with the use of data mining techniques and processes as a tool against public corruption, disclosing the potentiality of anomaly detection processes when evaluating public databases quality, on the one hand, and the discovery of information suspected of containing corrupt behavior, on the other.

The attributes that contributed the highest amount of entropy to the affidavit system, thus jeopardizing the quality of the database, were: property area, property value, shareholding percentage, property name, and year. Due to their low relative density, the property area is the most suspicious of containing anomaly data. In addition to revealing inconsistencies in the database, the possible detected anomalies could also show signs of corrupt behavior in relation to the fiscal value of the properties declared by the public officials since the characteristics of the property could affect its patrimonial value, according to whether it is declared at its fiscal or subfiscal value, or simply by avoiding its value declaration. In this regard, both the rules of behavior and the input-output bins could serve as a strategy of civic and accounting investigation in the fight against public corruption, and tax evasion and avoidance by public officials; all outrageous behaviors when considering the ideal exemplariness of the official in respect to the represented community.

Future research work could include the design of algorithmic variations in the proposed alphanumeric processes without discarding the application of variants to the alphanumeric processes proposed as well as to the processes of information mining external to the detection of anomalies and noise on the same affidavit databases.